\documentclass[runningheads]{svmult}
\usepackage{makeidx}   
\usepackage{graphicx}  
\usepackage{subeqnar}  
\usepackage{multicol}  
\usepackage{cropmark} 
\usepackage{subfigure}
\usepackage{physprbb}  
\def\msun{M_{\odot}}
\begin{document}

\title*{Dynamics of Dark-Matter Cusps}

\toctitle{Dynamics of Dark-Matter Cusps}
\titlerunning{Dark-Matter Cusps}
\author{David Merritt\inst{1}
\and Milo\v s Milosavljevi\' c\inst{1}}
\institute{Rutgers University, New Brunswick, NJ, USA}

\maketitle             

\begin{abstract}
Formation and disruption of dark-matter cusps are reviewed.
Accumulation of baryons at the center of a halo
can displace the dark matter,
converting singular density cusps into low-density cores.
The displaced mass can be of order $\sim 10M_{\bullet}$
with $M_{\bullet}$ the mass of the infalling population.
If $M_{\bullet}$ is identified with the masses of the
black holes currently observed at the centers of bright galaxies,
predicted core radii are $\sim {\rm a\ few} \times 10^2$ pc.
Other mechanisms, such as early mass outflow,
may explain the large dark-matter cores in dwarf and low-surface-brightness
galaxies.
Predictions of dark matter annihilation radiation from the center of 
the Milky Way galaxy are shown to be 
strongly dependent on the galaxy's merger history. 
\end{abstract}

\section{CDM Cusps}

The influential paper of Navarro, Frenk \& White \cite{NFW96}
proposed a universal, broken power-law profile for dark matter halos
in cold dark matter (CDM) cosmologies.
The dependence of $\rho$ on distance from the halo
center was found to be well described by a two-parameter family
of distributions,
$$
\rho_{NFW}(r) = \rho_0 \xi^{-1}\left(1+\xi\right)^{-2},\ \ \ \ \xi\equiv r/r_s.
$$
The central density of the NFW profile diverges as $\rho\sim r^{-1}$, 
similar to what was found in earlier 
$N$-body studies (e.g. \cite{DC91}).
However the small-radius dependence was little more than an
ansatz since the relevant scales were barely resolved in the simulations.
A debate ensued as to whether the profiles are indeed universal, 
and if so, what power of radius describes the dark matter density 
in the limit $r\rightarrow 0$.
A number of subsequent studies \cite{fum97,ghi98,moo98,moo99,kly01} 
found steeper central profiles,
e.g. $\rho\sim r^{-1.5}$, the ``Moore'' profile.

At about the same time, the luminous (stellar) densities in
galaxies were found to be centrally divergent
\cite{fer94,lau95}; in contrast with CDM halos, 
the power law index of stellar nuclei varies widely but correlates well with
galaxy properties, in the sense that fainter galaxies
have steeper cusps \cite{geb97}.
While the power-law nature of stellar cusps suggests a gravitational
origin, gas dynamics probably also played a role in their formation,
including processes like radiative dissipation, star formation,
and energy infusion from supernova explosions.

The power-law nature of density stratification in virialized
objects arises naturally in the standard
cosmological paradigm where the growth of objects proceeds from smallest
to largest scales \cite{prs74} starting from a featureless spectrum
of primordial fluctuations $P(k)\propto k^n$.  
Building blocks of structure can be identified
with peaks in the field of primordial Gaussian fluctuations.
The ensemble-averaged ``profile'' of an individual density peak is then a
power law with slope related to the tilt $n$
of the primordial spectrum on a given scale
\cite{hos85,syw98,lok00}. 
Furthermore, evolution of virialized systems through mergers
tends to preserve the power-law slope of the central cusp \cite{bar99}.  
Outside of the central cusp, the profile will be modified by the
tidal field of neighboring peaks and by the redistribution of material
in unequal-mass mergers.

It was noticed rather quickly following the advent of realistic
large-scale cosmological $N$-body simulations that after the slope
and normalization of the density profiles are taken into account,
cusps predicted by the simulations are too concentrated to accommodate
dynamical constraints derived from the rotation curves of (especially
low surface brightness) galaxies \cite{BGH96,GB98}.
Better fits were obtained with ad hoc, constant-central-density profiles,
e.g. the non-singular ``isothermal sphere'' \cite{dmr01,ils01,jlo02}.
Core sizes are not well constrained but recent model fits to the
rotation curve data give $r_c< 1$ kpc for most galaxies (e.g. \cite{jlo02}).
Other observations on galaxy scales (such as strong lensing 
\cite{rum01,kee01}) have led to more equivocal assessments of the
compatibility of dark matter models with observations.
Meanwhile, the most detailed study to date of resolution effects in the CDM
simulations \cite{pow02}
argues that the innermost resolved cusp slope is marginally steeper than
NFW, $\rho\sim r^{-1.2}$, and that it exhibits a shallowing trend with
decreasing radius.  

The cusp problem in CDM cosmologies can be alleviated either
by postulating changes in the basic physics, or by adding
baryons that couple to the dark matter and modify its profile.
Proposed changes in fundamental physics 
include weakly-interacting particles with 
large scattering cross-sections \cite{CMH92,SS00},
``warm'' dark matter \cite{SLD01},
broken scale-invariance \cite{KL99},
or modified gravity \cite{SM02,AH00}.
Standard-physics solutions include outflow of gas
during an early phase of galaxy formation 
\cite{NEF96},
torques from a barlike perturbation \cite{BGS01,WK01}
or infall of a population of condensed objects which 
disrupt the cusp \cite{MC01,ESH01}.

Current notions of galaxy formation are not precise enough
to allow testing of most of these ideas.
Here we focus on the last mechanism,
displacement of dark matter by condensed objects,
because the dynamics are straightforward and the
predicted level of the effect is quantifiable.
A lower limit on the displaced mass is set by the known
masses of compact objects at the centers of galaxies,
the supermassive black holes (SBHs) \cite{MF02},
some of which are known to have been present
since $z\sim 6$ \cite{Fan00}.
Regardless of their detailed formation history,
these black holes must have grown partly through the coalescence
of less-massive black holes during galaxy mergers,
displacing and ejecting dark (and baryonic) matter.
Predicted core sizes are $\sim$ hundreds of parsecs, 
consistent with the sizes of {\it stellar} cores in 
moderate- to high-luminosity galaxies, which probably
formed in the same way.
While it remains to be seen whether this model can explain
all of the discrepancies between predicted and observed
densities at the centers of halos, particularly in
low-surface-brightness and dwarf galaxies, 
the model is almost certainly relevant to the dark matter
distribution within a few tens of parsecs of the SBHs 
where dark matter annihilation radiation would be peaked.

\section{Enhancement and Disruption of Cusps}

Accumulation of baryons at the center of a dark-matter halo
will deepen the gravitational potential and steepen the cusp,
but the net effect depends on whether the baryons also
deposit energy into the dark matter as they fall in.
The simplest model (Fig. 1a) is adiabatic growth of a central point
mass due to radial infall of gas \cite{ips87,Peebles72}.
Let the initial mass of the central lump be zero and its final mass
$M_{\bullet}$.
For simplicity, assume that the orbits of the dark matter particles
are circles; if growth of the central lump is slow, 
the orbits will remain circular due to adiabatic invariance.
Conservation of angular momentum implies $r_iM_i=r_fM_{\bullet}$ where
$r_i$ and $r_f$ are the initial and final radii of a dark-matter
particle, $M_i$ is the dark mass within $r_i$ and $M_{\bullet}$ 
is assumed to dominate the potential within $r_f$.
Conservation of dark matter further implies $M_i(r_i)=M_f(r_f)$.

If the initial dark matter density profile is a power law,
$\rho_i\sim r^{-\gamma}$, the two conservation laws imply
$\rho_f\sim r^{-\delta}$, $\delta=(9-2\gamma)/(4-\gamma)$.
A more realistic, isotropic orbital distribution predicts the
same relation between $\gamma$ and $\delta$ \cite{GS99}.
For $0<\gamma<2$, the index of the power-law ``spike'' so 
formed lies between $2.25$ and $2.5$;
the spike extends roughly to the outer radius of influence
of the central object, $r\sim r_h\equiv GM_{\bullet}/\sigma^2$, 
where $\sigma$ is the velocity dispersion of the dark-matter 
particles in the pre-existing halo.
In physical units,
$$
r_h \approx 10 M_8\sigma_{200}^{-2} \ {\rm pc}
$$
where the central mass has been normalized to $10^8\msun$ and
the velocity dispersion to $200$ km s$^{-1}$.
This model has been widely used to predict the flux of annihilation
radiation from dark matter particles around the Milky Way black hole;
most of the radiation would come from the spike \cite{GS99,Gondolo00}.

\begin{figure}
\begin{center}
\includegraphics[width=.6\textwidth,angle=270.]{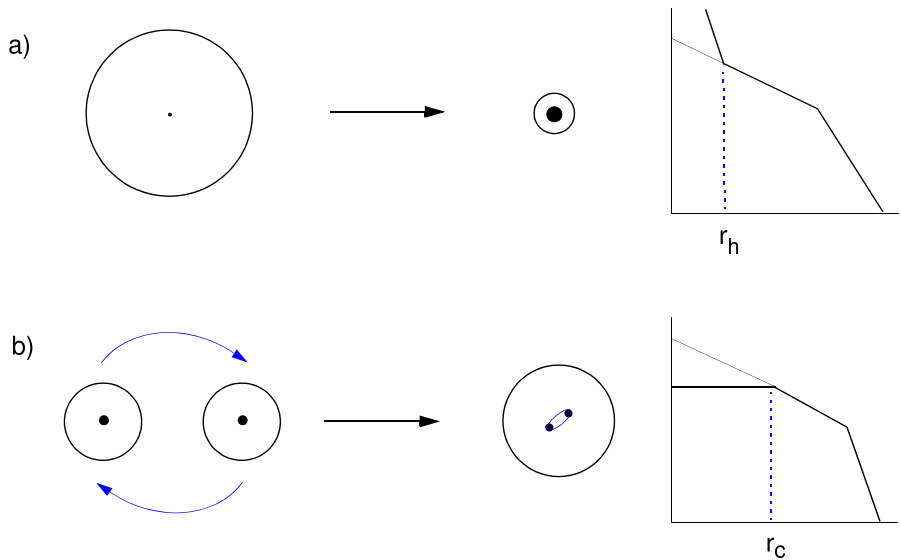}
\end{center}
\caption{
Two schemes for accumulating baryons at the center of a dark-matter
halo. (\textbf{a}) Mass falls in radially without depositing
 energy into the dark matter.
The central potential deepens and the cusp responds by becoming
steeper.
(\textbf{b}) 
Lumps form off-center and spiral in,
as they would following a merger.
As their orbits decay via dynamical friction, the lumps transfer
energy to the dark matter and lower its density.
}
\end{figure}

The outcome is very different if the lumps form
off-center, or are deposited in the halo following a merger
(Fig. 1b).
The lumps spiral inward due to dynamical friction, 
transferring their energy into the background and
{\it lowering} the central density
\cite{MC01,ESH01,Merritt83,NM99,UZK01}.
One well-studied case is the formation of a binary SBH
following a merger.
The binary first forms at a separation of $a\sim GM_{\bullet}/\sigma^2$,
with $M_{\bullet}\equiv m_1 + m_2$ the binary mass.
Continued evolution of the binary occurs through the 
{\bf gravitational slingshot} \cite{SVA74}: 
dark matter particles that approach the 
binary within a distance $\sim 3a$ are kicked out at
velocities of $\sim (GM_{\bullet}/a)^{1/2}$.

Slingshot ejection occurs at a rate determined by the coupled equations
\begin{eqnarray}
{d\over dt}\left({1\over a}\right) &=& H{G\rho\over\sigma}, \\
{dM_{ej}\over d\ln (1/a)} &=& J M_{\bullet}
\end{eqnarray}
\cite{MV92}.
Here $M_{ej}$ is the mass ejected by the binary and
$\rho$ and $\sigma$ refer to the background.
The dimensionless numbers $H$ and $J$ are functions of 
$m_1/m_2$ and the binary separartion; 
in the limit of a ``hard'' binary, $a\ll r_h$, and for $m_1\approx m_2$,
$H\sim 16$ and $J\sim 1$ \cite{Quinlan96}.
Solving the coupled equations for $M_{ej}(t)$ is not straightforward 
however since $\rho$ changes with time as the dark matter is ejected.
Simple models that take into account this time dependence \cite{Merritt00}  
give $M_{ej}\propto M_{\bullet}\ln t$,
consistent with what is seen in detailed
$N$-body simulations \cite{MM01}.

Fig. 2 shows the results of one such simulation \cite{MM01}.
The merger was between equal-mass spheroids with steep
central cusps, $\rho\sim r^{-2}$, and point particles 
representing the two SBHs.
The $r^{-2}$ cusp is gradually converted into an $r^{-1}$
cusp by the gravitational slingshot; at the end of the simulation, the
ejected mass is roughly twice the mass of the binary SBH.

\begin{figure}
\begin{center}
\includegraphics[width=1.\textwidth,angle=0.]{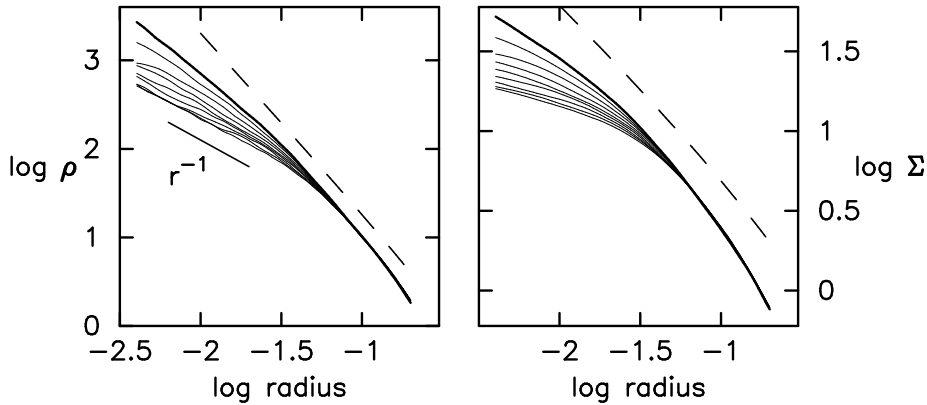}
\end{center}
\caption{Formation of a core by a binary SBH \cite{MM01}.
(a) Density profiles.
Dashed profile is the initial model, $\rho\sim r^{-2}$;
solid lines show the evolution of $\rho$ starting roughly
at the time when a hard binary is formed.
(b) Surface density profiles.
}
\end{figure}

The total mass ejected by a point-mass binary depends on how
long the binary continues to shrink.
An upper limit in the case of binary SBHs is given by the mass 
ejected in decaying to a separation of $a_{gr}$, 
the distance at which emission of gravitational radiation
induces a rapid coalescence.
The ejected mass is
\begin{equation}
M_{ej} \approx m_2\ln\left({a_h\over a_{gr}}\right)
\end{equation}
with $m_2$ the mass of the smaller black hole and
$a_h\sim GM_{\bullet}/\sigma^2$ \cite{Merritt00}.
For binary SBHs,
$M_{\bullet}\sim 10^8\msun$, $a_{gr}\sim 0.01$ pc
and the total mass ejected in decaying to $a_{gr}$ is 
$\sim 5M_{\bullet}$.
However a binary may ``stall'' after ejecting less mass
if the density in its vicinity drops too low \cite{MM01}.

If the number of massive infalling objects exceeds two,
heating of the background by binaries can occur indirectly as well:
massive particles are given kicks,
then lose their kinetic energy to the background
as they spiral in and are kicked out again etc.
As a crude model of this process, 
consider the evolution of an ensemble of point particles
of mass $m_{\bullet}$ orbiting in a dark-matter halo
with $\rho\sim r^{-2}$ -- slightly steeper than the cusps found
in CDM simulations but easier to treat.
The energy released as one lump spirals in from radius $r_i$ to $r_f$
is $2m_{\bullet}\sigma^2\ln(r_i/r_f)$.
Decay will halt when the lumps form a self-gravitating system
of radius $\sim GM_{\bullet}/\sigma^2$ with $M_{\bullet}$ 
the total baryonic mass.
Equating the energy released during infall with the energy 
of dark matter initially within
$r_c$, the ``core radius,'' gives
\begin{equation}
r_c \approx {2GM_{\bullet}\over\sigma^2}\ln\left({r_i\sigma^2\over GM_{\bullet}}\right).
\end{equation}
Most of the particles that deposit their energy within $r_c$ 
will come from radii $r_i \approx {\rm a\ few} \times r_c$,
implying $r_c\approx {\rm \ several} \times GM_{\bullet}/\sigma^2$
and a displaced mass of $\sim {\rm \ several}\times M_{\bullet}$.

Evolution will continue as the massive particles form binaries and 
begin to kick out other massive particles.
Assume that ejection occurs via many small kicks, such
that almost all of the binding energy so released can
find its way into the dark matter as the ejected particles
spiral back into the core.
The energy released by a single binary in shrinking
to a separation such that its orbital velocity equals
the escape velocity from the core is
$\sim m_{\bullet}\sigma^2\ln(4M_{\bullet}/M_{halo})$.
If all of the massive particles find themselves in such
binaries before their final ejection and if most of their
energy is deposited near the center of the halo, the
core mass will be
\begin{equation}
M_{ej} \approx M_{\bullet} \ln \left({M_{halo}\over M_{\bullet}}\right)
\end{equation}
e. g. $\sim 10 M_{\bullet}$ for $M_{\bullet}/M_{halo}\approx 10^{-4}$.

\begin{figure}
\begin{center}
\includegraphics[width=.7\textwidth,angle=-90.]{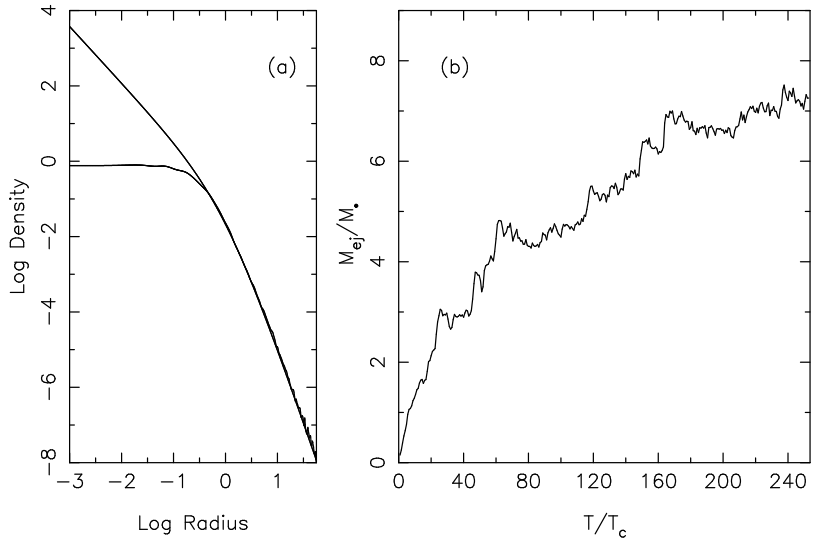}
\end{center}
\caption{Formation of a core by infalling lumps; details of
the simulation are given in the text.
(a) Initial and final density profiles of the dark matter.
(b) Ejected dark mass as a function of time,
defined as the difference in enclosed mass between the 
 profiles at times $0$ and $T$ inward of their first intersection.
$M_{\bullet}$ is the total ``baryon'' mass, i.e. the mass
of the infalling population.
$T_c$ is the crossing time.
Almost all of the massive objects have been ejected by the 
end of the simulation.
}
\end{figure}

A detailed $N$-body simulation illustrates this process
(Fig. 3).
The initial model for the dark-matter halo has 
$\rho\sim r^{-1.5}$ at small radii and total mass $M$.
The condensed objects were represented by 20 point masses,
each weighing 10 times as much as a dark matter particle
and with combined mass $M_{\bullet}=10^{-2}M$.
Evolution occurs in two stages, as discussed above.
As the orbits of the massive particles initially decay,
they transfer energy to the background 
and lower the central density.
Once the massive particles form a bound subsystem,
further evolution is driven by the formation and disruption
of binaries.
The result is a slower growth of the dark matter core, as massive
particles are ejected from the central regions and fall
back in.
At the end of the simulation the dark mass ejected from the cusp 
is $\sim 8 M_{\bullet}$.

Only a single massive binary is left at the end -- most of the particles that 
displaced the dark matter have been ejected from the system.
This model is the dynamical analog of wind-driven
outflow models \cite{NEF96,BGS01}:
in both schemes, the baryons that created the core
are no longer in evidence after the core is in place.

\begin{figure}
\begin{center}
\includegraphics[width=1.\textwidth]{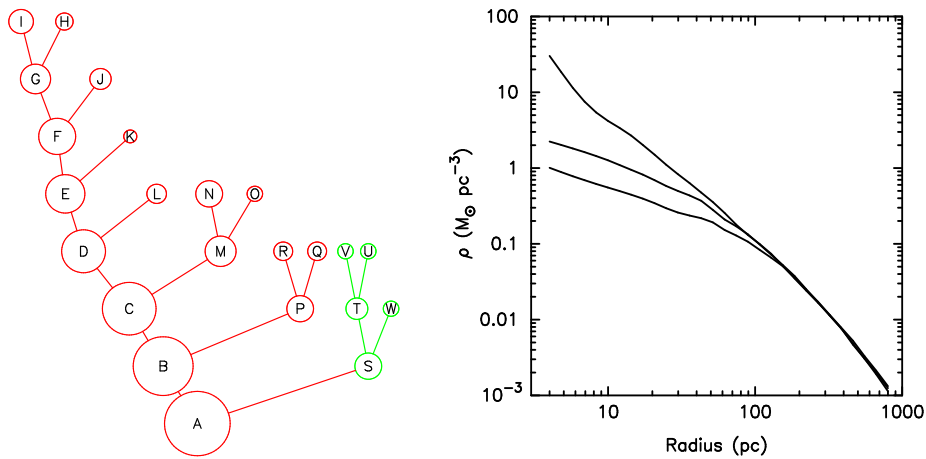}
\end{center}
\caption{
Cusp destruction in hierarchical mergers of halos containing
central ``black holes'' \cite{MM02}.
Left panel: a merger tree; time proceeds downwards and the area
of each circle is proportional to the mass of the halo, and
of its central SBH.
Right panel:
Halo profiles.
The top profile is the initial halo model (e.g. U, V);
the middle profile is halo T; the bottom profile is
halo S.
The profiles are scaled so that the black hole mass is
the same in each.
The progenitor halos were truncated Moore profiles,
with $\rho\sim r^{-1.5}$ central density cusps.
Axis scalings are based on an assumed black hole mass
of $10^8\msun$, a halo mass of $10^{13}\msun$ and a virial
radius of $300$ kpc.
}
\end{figure}

Is this a realistic model?
Multiple baryonic clumps are often postulated to form via
gaseous dissipation in the early stages of galaxy formation
(e.g. \cite{BGS01,ESH01}) but their masses, numbers and degree
of central concentration are poorly known.
Another possible candidate for the lumps is an early population
of black holes, e.g. the remnants of Population III stars
($m_{\bullet}\sim 150\msun$) \cite{BCL99}.
But current theory suggests that only one such black hole forms per 
mini-halo at a redshift of $\sim 20$ \cite{ABN02}, and furthermore
the inspiral time for an object of this mass would be of order
$10^9$ yr, unless it came attached to a larger mass.
Black holes of intermediate mass ($m_{\bullet}\sim 10^3\msun$)
are a third possibility but their existence is debated and
their time of formation uncertain.

SBHs have been present in at least
some spheroids since a redshift of $\sim 6$ \cite{Fan00};
many of the SBHs that we see now should therefore
be the products of multiple mergers.
Typically those mergers will have been ``minor,'' $m_2\ll m_1$.
In such an interaction the ejected mass scales with $m_2$,
the mass of the smaller hole; the total mass ejected
in a sequence of such mergers  
would then depend only on $M_{\bullet}$, the final mass.
However if a black hole grows via a merger
hierarchy of comparably-massive black holes,
the effect on the density cusp should be to a certain extent 
{\bf cumulative}: a merger of two galaxies
whose cusps had previously been destroyed by binary black holes,
will produce a shallower central profile than a merger between two
galaxies with initially steep cusps.
This is verified via an $N$-body experiment in Fig. 4.
The ejected mass scales both with
$M_{\bullet}$, the final black hole mass, and $N$, the number
of stages in the merger hierarchy.

There is some quantitative empirical support for idea that
cores are produced from the binding energy released by 
binary black holes.
The luminous (stellar) matter in elliptical galaxies 
and bulges exhibits steep power-law cusps in faint
galaxies and cores in bright galaxies \cite{geb97};
among the ``core'' galaxies, the core mass is 
of order $\sim 10M_{\bullet}$ with $M_{\bullet}$ the 
current black hole mass \cite{MMRV02}.
The largest cores observed, in galaxies like M87,
have sizes of $\sim 0.5$ kpc; it seems circumstantially reasonable
to suppose that dark-matter cores could be comparable
in size.
The persistence of steep cusps in faint elliptical galaxies is puzzling
but may indicate that the stellar cusps formed by dissipative processes
during or after the last significant merger \cite{SM96};
in any case, presence of a stellar cusp is not necessarily evidence for
a dark matter cusp.

SBHs represent an extreme in terms of
compactness, but galaxies contain other condensed components
that could in principle act like ``black holes'' during mergers, 
displacing dark matter as they spiral inward.
Bulges are an obvious candidate; with masses $\sim 10^3$
times those of SBHs \cite{MF01} they could in principle
displace a much larger quantity of dark matter.
In order for two bulges to act like a ``binary black hole,''
they must be compact enough to not overlap appreciably at
the separation where they first form a bound subsystem.
This separation is $\sim GM_{bulge}/\sigma^2$ with $\sigma$
the dark matter velocity dispersion.
Requiring this separation to be greater than the bulge
scale length $R_{bulge}$ gives $GM_{bulge}\gg R_{bulge}\sigma^2$
or $\sigma_{bulge}\gg\sigma$, with $\sigma_{bulge}$ the bulge
velocity dispersion.
Thus a ``hot'' bulge is a ``hard'' bulge.
In fact $\sigma_{bulge}$ appears to be comparable to,
though somewhat larger than, $\sigma$ \cite{Ferrarese02}
so it is unlikely that bulges -- at least as we currently
observe them -- would be effective at displacing dark matter
during mergers.

\section{An Application: Annihilation Radiation from the Galactic Center}

If dark matter consists of massive weakly interacting particles
like neutralinos, the particles should self-annihilate producing
a potentially observable signal.
The annihilation rate in a spherical halo is
\begin{equation}
\Gamma={\sigma v\over M_{\chi}^2}\int \rho^2(r) 4\pi r^2\ dr,
\end{equation}
with $M_{\chi}$ the neutralino mass, $\sigma v$ the annihilation
cross section, and $\rho$ the dark matter density \cite{BUB98}.
Annihilation products include gamma rays, from the decay of neutral
pions; high-energy neutrinos; and synchrotron emission
from high energy positrons and electrons moving in the ambient
magnetic field.
Since the annihilation flux depends on the squared density of neutralinos, 
the signal would be greatly enhanced along lines of sight where 
the dark matter is clumped.
These include the center of the Milky Way galaxy, where the density
in a smooth halo would be maximum, as well as directions 
that intersect the centers of relic halos orbiting as subclumps in
the Milky Way halo \cite{BEGU99,CRM99}.

\begin{figure}
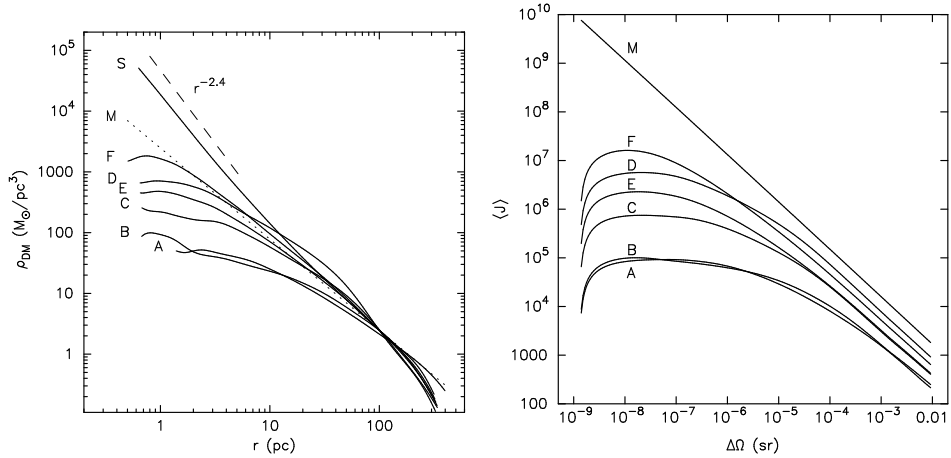

\begin{center}
\mbox{\subfigure
{\includegraphics[width=0.5\textwidth]{fig5a.ps}}\quad
\subfigure
{\includegraphics[width=0.5\textwidth]{fig5b.ps}}
}
\end{center}
\caption{Dark matter annihilation radiation from the galactic
center \cite{MMVJ02}.
Left panel: Density profile of the Milky Way dark halo.
Curves labelled M and S show the halo profile
before and after growth of the central SBH.
Other curves show profiles following mergers with smaller
halos with various mass ratios: (A, B) 1:1; (C, D) 1:3; E 1:5; F 1:10.
The mass of the SBH in the smaller halo scales with halo mass.
Right panel: Dimensionless integrated annihilation flux as a function of 
the angular acceptance of the detector, centered on the SBH.}
\end{figure}

The annihilation signal from the galactic center would be further enhanced
if there is a power-law spike associated with the central SBH,
as discussed above.
Gondolo \& Silk \cite{GS99} considered this case and showed that the flux
is increased by $\sim 5$ orders of magnitude compared with
models lacking a central spike.
This prediction has been claimed to be inconsistent with measured fluxes
of annihilation products, including synchrotron emission \cite{Gondolo00}
and neutrinos \cite{Auriemma02} (but see \cite{BSS01,BSS02}).

The central slope of the Milky Way's dark matter distribution could 
have been substantially lowered if the halo experienced a merger
after formation of the spike, assuming that the infalling halo
also contained a SBH.
At least one such merger, with a mass ratio of $1:10$ or less,
is statistically likely to have been experienced by a halo
as massive as that of the Milky Way since a redshift of $2$
\cite{MMVJ02}.
Fig. 4 shows density profiles of merged dark matter halos,
in simulations where the initial profile of the larger halo was
normalized to match the Milky Way, with a central $r^{-1.5}$ density
cusp \cite{MMVJ02}.
The spike is efficiently destroyed by the binary 
SBH in all of the mergers, greatly reduced the predicted
annihilation flux.

Cusp destruction may have been avoided in the case of our galaxy
\cite{BSS02}.
Little is known about the detailed merger history of the Milky Way
but it has been argued that the ages of stars in the galaxy's
``thick disk'' imply a time of $\sim 11$ Gyr since the last significant
merger \cite{Wyse01}.
The Milky Way SBH may have acquired most of its mass during or
even after this merger event.
However, if the SBH grew significantly after a merger that had destroyed a 
pre-existing dark matter spike, it is unlikely that the new
spike would be as steep.
The reason is that the dark matter density would probably 
be non-singular following the merger (e.g. Fig. 4), and growth of a point
mass in a non-singular background produces a shallow spike,
$\rho\sim r^{-1.5}$ \cite{Peebles72}.
The annihilation flux from such a spike would be only
slightly greater than the expected background in the direction
of the galactic center \cite{GS99}.

\bigskip
\noindent {\bf Acknowledgments}

The $N$-body simulation illustrated in Fig. 3 was carried out by 
M. Hemsendorf, who kindly gave us permission to show his results here.
The merger tree shown in Fig. 4 was generated from a computer code
written by L. Verde.
We thank R. Jimenez, C. Power and L. Verde for comments
on the manuscript.
This work was supported by NSF grant AST 00-71099 and NASA grants
NAG5-6037 and NAG5-9046.

\end{document}